\documentclass[12pt]{article}
\usepackage{color}
\usepackage{amsfonts,amssymb}
\usepackage{epsfig, picinpar}
\usepackage{amssymb,amscd}

\newtheorem{theorem}{Theorem}

\newtheorem{remark}{Remark}

\newtheorem{corollary}{Corollary}

\newcommand{\zhen}[9]{\mbox{$
  \left(\begin{array}{ccc}{#1} & {#2} & {#3}\\{#4} & {#5} & {#6}\\
{#7} &{#8} &{#9}  \end{array}\right)$}}
\newcommand{\plem}[3]{\mbox{$\left(\begin{array}{c} \psi_{#1}\\ \psi_{#2}\\ \psi_{#3}   \end{array} \right)$}}
\newcommand{\phlem}[3]{\mbox{$\left(\begin{array}{c} \psi^*_{#1}\\ \psi^*_{#2}\\ \psi^*_{#3}   \end{array} \right)$}}

\newcommand{\R}{\mathbb{R}}

\newcommand{\Z}{\mathbb{Z}}

\newcommand{\lam}{\lambda}

\newcommand{\pa}{\partial}

\newcommand{\al}{\alpha}  
\newcommand{\be}{\begin{equation}}
\newcommand{\ee}{\end{equation}}
\newcommand{\beq}{\begin{eqnarray}}
\newcommand{\eeq}{\end{eqnarray}}
\newcommand{\beqq}{\begin{eqnarray*}}
\newcommand{\eeqq}{\end{eqnarray*}}

\newcommand{\La}{\Lambda}

\title{\bf An integrable hierarchy, parametric solution
and traveling wave solution}
\author{Darryl D. Holm, \ Zhijun Qiao\\
T-7 and CNLS, MS B-284, Los Alamos National Laboratory\\
Los Alamos, NM 87545, USA\\
{\footnotesize  E-mails:dholm@lanl.gov \ qiao@lanl.gov}
}
\date{1st version April 10, 2002; this version Oct. 30, 2002}
\begin{document}
\maketitle

\begin{abstract}
This paper gives an integrable hierarchy of nonlinear evolution
equations. In this hierarchy there are  
the following representative  equations:
\beqq
& & u_t=\pa^5_x u^{-\frac{2}{3}},\\
& &  u_t=\pa^5_x\frac{(u^{-\frac{1}{3}})_{xx}
    -2(u^{-\frac{1}{6}})_{x}^2}{u};\\
& & u_{xxt}+3u_{xx}u_x+u_{xxx}u=0.
\eeqq
The first two are in the positive order
hierarchy while the 3rd one is in the negative order hierarchy.
The whole hierarchy is shown integrable through
solving a key $3\times 3 $ matrix equation. 
The $3\times3$ Lax pairs
  and
their adjoint representations  are nonlinearized to be two
Liouville-integrable canonical Hamiltonian systems. 
Based on the integrability
of $6N$-dimensional systems we give
the parametric solution of the positive
hierarchy. In particular, we obtain the parametric solution 
of the equation  $u_t=\pa^5_x u^{-\frac{2}{3}}$.
Moreover, we give the traveling
wave solution (TWS) of the above three equations. The TWSs of the first two
equations have singularity and look like cusp (cusp-like), but the TWS of the 3rd one is continuous.
For the 5th-order equation, its   parametric solution
can not include its singular TWS. 
We also analyse the Gaussian initial solutions for the equations
$u_t=\pa^5_x u^{-\frac{2}{3}}$, and  $u_{xxt}+3u_{xx}u_x+u_{xxx}u=0.$
One is stable, the other not. Finally, we extend the equation
$u_t=\pa^5_x u^{-\frac{2}{3}}$ to a large class
of equations $
u_t=\partial_x^l u^{-m/n}, \  \ l\ge1,\ n\not=0,\ \ m,n \in \Z,
 $ which still have the singular cusp-like traveling wave solutions.
  
 \end{abstract}
   
   $ $\\
{\it AMS Subject: 35Q53; 58F07; 35Q35}

\section{Introduction}

The inverse scattering transformation (IST) method plays a very
important role in the investigation of integrable nonlinear evolution equations (NLEEs) \cite{GGKM}.
  This method has been successfully applied
to solve the integrable NLEEs in the form of soliton solutions.
These NLEEs include the well-known KdV
equation \cite{KDV}, which is related to a 2nd order operator
(i.e. Hill operator) 
spectral problem \cite{LG, Mar}, the remarkable AKNS equations
\cite{AKNS, AKNS1}, which is associated with the Zakharov-Shabat (ZS)
spectral problem \cite{ZS}, and other higher dimensional
integrable equations.

In the theory of
integrable system, it is significant for us to search for as many new
integrable evolution equations as possible. Kaup \cite{Kaup1} studied
the inverse scattering problem for cubic eigenvalue equations of the
form $\psi_{xxx}+6Q\psi_x+6R\psi=\lam\psi$, and showed a 5th
order partial differential equation (PDE)
 $Q_t+Q_{xxxxx}+30(Q_{xxx}Q+\frac{5}{2}Q_{xx}Q_x)
+180Q_xQ^2=0$ (called the KK equation) integrable. Afterwards, Kuperschmidt \cite{Kup1}
constructed a super-KdV equation and presented the integrability of
the equation through giving bi-Hamiltonian property and Lax
form. Very recently, Degasperis and Procesi \cite{DP[1999]}
proposed a new integrable equation (called DP equation) with the soliton solution of peakon
type.  The DP equation is an extension of the Camassa-Holm (CH) equation
 \cite{CH1},  and is proven to be associated with
a 3rd order spectral problem \cite{DHH[2002]}:
$\psi_{xxx}=\psi_x-\lam m\psi$ and to have some
relationship to a canonical Hamiltonian system
under a new nonlinear  Poisson bracket (called Peakon Bracket) 
 \cite{HH[2002]}. In \cite{Holm-Qiao-2002}, we extended the DP equation to
an integrable hierarchy and deal with its parametric solution
and peaked stationary solutions.

In Ref. \cite{DHH[2002]}, the authors
studied the DP equation, of a similar form to
the Camassa-Holm shallow water wave equation, 
and proved the exact integrability of this equation by
constructing its Lax pair. The DP equation is related to  a
negative  flow in the Kaup-Kupershmidt hierarchy via a reciprocal
transformation. The infinite sequence of conserved quantities is
derived together with a proposed bi-Hamiltonian structure. The
equation admits exact solutions in the form of a superposition of
multi-peakons, and  is compared with the analogous results for Camassa-Holm
peakons.

The present work  is motivated on the basis of the eigenvalue problem
$ \psi_x-\al^2\psi_{xxx}=\al^2\lam m\psi$, which was introduced by
Desgaperis, Holm and Hone \cite{DHH[2002]}. Here, we consider the limit
case
of $\al $ going to infinity. That is, we get  the 3rd order
spectral problem $\psi_{xxx}=-\lam m\psi $.
In this paper, starting from that problem, we give an integrable hierarchy, and
through solving a key matrix equation we explicitly
provide the Lax pairs for the whole
hierarchy.  
The following equations
\beq
& & u_t=\pa^5_x u^{-\frac{2}{3}},\label{newequ}\\
& &  u_t=\pa^5_x\frac{(u^{-\frac{1}{3}})_{xx}
    -2(u^{-\frac{1}{6}})_{x}^2}{u}; \label{newequ1}\\
& & u_{xxt}+3u_{xx}u_x+u_{xxx}u=0,\label{newequ2}
\eeq
are three representatives in the hierarchy. 
 The first equation is a reduction of some $2+1$ dimensional equation
\cite{KD1}.
The second one is new, and the third one is actually 
twice derivatives in $x$ of the Rieman shock equation 
\cite{DHH[2002]}.

 Konopelchenko and Dubrovsky \cite{KD1}
already pointed out  equation (\ref{newequ}) is integrable
from the reduction view point 
\cite{KD1}, but  did not discuss  the spectral problem
and the solution of the equation.
 Here we deal with its spectral problem
and representation of solution 
from the constraint
view point. We give the parametric solutions
for the hierarchy, particularly  
 for equation (\ref{newequ}).
Furthermore, we obtain the 
traveling wave solution (TWS) for equations
(\ref{newequ}), (\ref{newequ1}), and (\ref{newequ2}).
The first two look like
a class of  cusp  soliton solutions (called `cusp-like', but not cusp soliton
\cite{Wadati}).
The TWSs of 
equations (\ref{newequ}) and (\ref{newequ1}) have singularity, 
but the TWS of  equation (\ref{newequ2}) is continuous.
Additionally, for the 5th-order equation (\ref{newequ}), its smooth  parametric solution
can not include its singular TWS. 
Equation (\ref{newequ2}) has the compacton-like  and parabolic cylinder solutions. 
We also analyse the Gaussian initial  solutions for equations
$u_t=\pa^5_x u^{-\frac{2}{3}}$ and  $u_{xxt}+3u_{xx}u_x+u_{xxx}u=0.$
The former is stable, and the latter not (see Figures 7, 1 - 5).
 Finally, we extend the equation
$u_t=\pa^5_x u^{-\frac{2}{3}}$ to a large class
of equations $
u_t=\partial_x^l u^{-m/n}, \  \ l\ge1,\ n\not=0,\ \ m,n \in \Z,
 $ which still have the singular cusp-like traveling wave solutions.  

The whole paper is organized as follows.
Next section is saying how to connect the above three 
equations 
to a spectral problem and how to cast them into a new hierarchy of
NLEEs. In section 3, we construct
the zero curvature representations for this new hierarchy
through solving a key $3\times 3 $ matrix equation. In particular, we
obtain the Lax pair of equations (\ref{newequ}), (\ref{newequ1}), (\ref{newequ2}), and therefore they
are  integrable.  In section 4, we show that the 3rd order spectral
problem related to the above three equations  is nonlinearized as a completely integrable
Hamiltonian 
system under some constraint in $\R^{6N}$. In section 5 we give the
parametric solution for the positive order hierarchy of NLEEs. 
We particularly get the parametric solution
of equation (\ref{newequ}). Moreover, in section 6
we obtain
the traveling wave solutions for equations (\ref{newequ}), 
(\ref{newequ1}), and  (\ref{newequ2}), and we also
analyse the  Gaussian initial solutions for the equations
$u_t=\pa^5_x u^{-\frac{2}{3}}$ and
$u_{xxt}+3u_{xx}u_x+u_{xxx}u=0$.  We propose a family
of new equations $
u_t=\partial_x^l u^{-m/n}, \  \ l\ge1,\ n\not=0,\ \ m,n \in \Z,
 $ which still have the singular cusp-like traveling wave solutions.
Finally, in section 7 
 we give some conclusions.

\section{Spectral problems and a new hierarchy}
Let us consider the following 3rd order spectral problem

\be
\psi_{xxx}=-\lam u\psi \label{sp1}
\ee
and its adjoint problem
\be
\psi^*_{xxx}=\lam u\psi^*. \label{sp2}
\ee

Take $u\rightarrow u+\epsilon \delta u $,  $\lam \rightarrow
\lam+\epsilon \delta \lam  $, and $\frac{\pa}{\pa
  \epsilon}|_{\epsilon=0} $.
Then, we have their functional gradient $\frac{\delta
  \lam}{\delta u}$ with respect to the potential $u$
\be
\frac{\delta
  \lam}{\delta u}=\frac{\lam\psi\psi^*}{E}\equiv\frac{\nabla\lam}{E}, \label{fg1}
\ee
where
\beq
\nabla\lam&=&\lam\psi\psi^*, \label{lpp}\\
E&=&\int_{\Omega}u\psi\psi^* dx=constant, \nonumber 
\eeq
and $\Omega=(-\infty, \infty)$ or $\Omega=(0, T)$. In this procedure, 
 we need the boundary conditions of $u$ decaying at infinities or
of $u$ being periodic with period $T$. Usually, we compute the functional gradient $\frac{\delta
  \lam}{\delta u}$ of the eigenvalue $\lam$ with respect to the
  potential $u$ by using the method in Refs. 
\cite{C1,TuGZ}.

Through doing five times derivatives of Eq. (\ref{lpp}),
we find
\beqq
(\nabla\lam)_{xxxxx}&=&-3\lam^2(2u\pa+\pa u)(\psi\psi_x^*-\psi^*\psi_x),\\
(\psi\psi_x^*-\psi^*\psi_x)_{xxx}&=&(u\pa+2\pa u)\nabla\lam,
\eeqq
which directly lead to 
\beq
K\nabla\lam&=&\lam^2J\nabla\lam, \label{KJ2}
\eeq
where
\beq
K&=&\pa^5,   \label{K2}\\
J&=&-3(2u\pa+\pa u)\pa^{-3}(u\pa+2\pa u). \label{J2}
\eeq

{\bf Hint: } \ Here we do not care about the Hamiltonian properties
of the operators $K, J$, but need
\beqq
K^{-1}&=&\pa^{-5},\\
J^{-1}&=&-\frac{1}{27}u^{-2/3}\pa^{-1}u^{-1/3}\pa^3u^{-1/3}\pa^{-1}u^{-2/3}.
\eeqq
They yield 
\beq
& &{\cal L}=J^{-1}K=-\frac{1}{27}u^{-2/3}\pa^{-1}u^{-1/3}\pa^3u^{-1/3}\pa^{-1}u^{-2/3}\pa^{5},
              \label{6.4.4}\\
& &{\cal L}^{-1}=K^{-1}J=-3 \pa^{-5}(2u\pa+\pa u)\pa^{-3}(u\pa+2\pa u).\eeq

By this  pair of  operators 
we define the hierarchy of nonlinear evolution equations
associated with the spectral problems (\ref{sp1}) and (\ref{sp2}). 
Let $G_0\in Ker\ J=\{G\in C^{\infty}(\R)\ |\ JG=0\}$ and
$G_{-1}\in Ker\ K=\{G\in C^{\infty}(\R)\ |\ KG=0\}$. We define
the Lenard sequence
\beq
G_j= \left\{\begin{array}{ll}
{\cal L}^j\cdot G_0, & j\ge 0,\ j\in\Z\\
{\cal L}^{j+1}\cdot G_{-1},& j<0, \ j\in \Z.\end{array}\right. 
\label{Gj}
\eeq
where ${\cal L}=J^{-1}K$ is called the recursion operator.
Therefore  we produce a new
 hierarchy of nonlinear evolution equations (NLEEs): 
\beq
u_{t_k}=J G_k, \ \forall k \in \Z. \label{mtk}
\eeq

Apparently, this hierarchy includes the positive order $(k\ge0)$
and the negative order $(k<0)$ cases.
Let us now give several representative  equations in the hierarchy (\ref{mtk}).

\begin{itemize}
\item  Choosing $G_{-1}=\frac{1}{6}\in Ker \ K$ yields 
       the first equation in the negative hierarchy:
  \beq
u_t+vu_x+3v_xu=0, \ u=v_{xx}. \label{m-1}
\eeq
This equation is actually:
$v_{xxt}+3v_{xx}v_x+v_{xxx}v=0$ which is equivalent to
$\pa^2(v_{t}+vv_{x})=0$.
Obviously, $v=c_1x+c_0$ ($c_1, \ c_0 $ are two constants) is a special
solution of this equation. In section 6, we will study
its traveling wave solution.

The second equation in the negative hierarchy is:
\beqq
& & u_{t_{-2}}+3(u_xw+3uw_x)=0,\\
& & w_{xxx}+\frac{9}{2}(2u_xv+3uv_x)=0,\\
& & u=3(\sqrt{v_{xx}})_{xx}.
\eeqq

\item Choosing $G_0=u^{-\frac{2}{3}}\in Ker \ J$ leads to  
       the second  equation in the positive hierarchy:
  \beq
u_t=\pa^5_x u^{-\frac{2}{3}}. \label{m+2}
\eeq
Konopelchenko and Dubrovsky
ever pointed out
that this equation is integrable and is a reduction of some $2+1$ dimensional equation
\cite{KD1}. But they did not study the solution
of this equation.
In the following, we study the relation between 
 this equation and finite-dimensional integrable system
and will find that it has parametric solution as well as
the traveling wave solution which looks like a cusp.

The third equation in the positive hierarchy is:
\beqq
& & u_{t_{2}}=-\frac{1}{27}(u^{-2/3}v)_{5x},\\
& & v_{x}=u^{-1/3}(u^{-1/3}w)_{3x},\\
& &w_x=u^{-2/3}(u^{-2/3}w)_{5x}.
\eeqq

\item
  Choosing another element $G_0=\frac{(u^{-\frac{1}{3}})_{xx}
    -2(u^{-\frac{1}{6}})_{x}^2}{u}\in Ker \ J$ gives  
       the following representative  equation in the positive hierarchy:
  \beq
u_t=\pa^5_x\frac{(u^{-\frac{1}{3}})_{xx}
    -2(u^{-\frac{1}{6}})_{x}^2}{u}. \label{m2}
\eeq
This equation also has a singular cusp-like traveling wave solution.

\end{itemize}
Of course, 
we may produce further nonlinear equations
by selecting  other elements from the kernels of $J, K$. In the following,
we will see that all equations in the hierarchy (\ref{mtk}) are integrable.
Particularly, {\bf the above three equations (\ref{m-1}), (\ref{m+2}),
(\ref{m2}) are integrable}.

\section{Zero curvature representations}

Letting $\psi=\psi_1 $, we change Eq. (\ref{sp1}) to a $3\times 3 $ matrix spectral problem
\beq
\Psi_x&=&U(u,\lam)\Psi,  \label{SP1}\\
U(u,\lam)&=&\zhen{0}{1}{0}{0}{0}{1}{-\lam u}{0}{0},\ \
\Psi=\plem{1}{2}{3}. \label{U1}
\eeq

Apparently, the Gateaux derivative matrix  $U_{*}(\xi)$ of the
spectral matrix  $U$  in the direction
$\xi\in C^{\infty}(\R)$ at point $u$ is
\be
U_{*}(\xi)\stackrel{\triangle}{=}\left.\frac{{\rm d}}{{\rm d} \epsilon}
\right|_{\epsilon=0}U(u+\epsilon\xi)
=\zhen{0}{0}{0}{0}{0}{0}{-\lam\xi}{0}{0} \label{6.4.5}
\ee
which is obviously an injective homomorphism, i.e.
$U_*(\xi)=0\Leftrightarrow \xi=0 $.

   For any given $C^{\infty}$-function $G$, we construct
    the following $3\times 3 $ matrix  equation with respect to
    $V=V(G)$
    \beq   V_x- [U,V]=U_{*}(K G-\lam^2 J G). \label{VLCH}
    \eeq

  \begin{theorem}
For the spectral problem (\ref{SP1}) and an arbitrary
$C^{\infty}$-function $G$, the matrix equation
(\ref{VLCH})  has the
following solution
{\normalsize \beq
  V=\lam
\zhen{-G''-3\lam\pa^{-2}\Upsilon G}{3(G'+\lam\pa^{-3}\Upsilon G)}{-6G}
{-G'''-3\lam\pa^{-1}uG'}{2G''}
{3(-G'+\lam\pa^{-3}\Upsilon G)}
{-G''''-3\lam^2u\pa ^{-3}\Upsilon G}
{G'''-3\lam\pa^{-1}uG'}{-G''+3\lam\pa^{-2}\Upsilon G},
 \label{6.4.15}
\eeq}
where $\pa=\pa_x=\frac{\pa}{\pa x}, \ \Upsilon=u\pa+2\pa u$, and the
supscript
$'$ means the derivative in  $x$. Therefore,
$J=-3\Upsilon^*\pa^{-3}\Upsilon $ ($\Upsilon^*$ is the conjugate of $\Upsilon$).
\label{Th1}
\end{theorem}
{\bf Proof:} \ \ Set 
\beqq
V=\zhen{V_{11}}{V_{12}}{V_{13}}{V_{21}}{V_{22}}{V_{23}}{V_{31}}{V_{32}}{V_{33}},
\eeqq 
and subsitute this into Eq. (\ref{VLCH}). This is a overdetermined
equation. Using some calculation techniques \cite{Qiao}, we obtain the following
results:
\beqq
V_{11}&=&-\lam G''-3\lam^2\pa^{-2}\Upsilon G,\\
V_{12}&=&3(\lam G'+\lam^2\pa^{-3}\Upsilon G),\\
V_{13}&=& -6\lam G,\\
V_{21}&=&-\lam G'''-3\lam^2\pa^{-1}uG', \\
V_{22}&=& 2\lam G'',\\
V_{23}&=&3\lam(-G'+\lam\pa^{-3}\Upsilon G),\\
V_{31}&=&-\lam G''''-3\lam^3 u\pa ^{-3}\Upsilon G,\\
V_{32}&=&\lam G'''-3\lam^2\pa^{-1}uG',\\
V_{33}&=&-\lam G''+3\lam^2\pa^{-2}\Upsilon G ,
\eeqq
which completes the proof.

\begin{theorem}
Let $G_0\in Ker\ J$, 
$G_{-1}\in Ker\ K$, and let each $G_j $ be given 
through Eq. (\ref{Gj}).
Then, 
\begin{enumerate}
\item 
  each new vector field $X_k=J G_k, \ k\in \Z$ satisfies the
following commutator representation
\beq 
V_{k,x}-[U,V_k]=U_*(X_k), \ \forall k\in \Z; \label{UV1}
\eeq
\item  the new  hierarchy (\ref{mtk}), i.e.
\beq
u_{t_k}= X_k=JG_k, \ \forall k \in \Z, \label{mtk1}
\eeq
possesses the zero curvature representation
\beq
U_{t_k}-V_{k,x}+[U,V_k]=0,\ \forall k\in \Z, \label{XUV}
\eeq
\end{enumerate}
where 
\beq
V_k&=&\sum V(G_j)\lam^{2(k-j-1)}, \ \ \sum\ = \
\left\{\begin{array}{ll}
\sum^{k-1}_{j=0}, & k>0,\\
0, & k=0,\\
-\sum^{-1}_{j=k}, & k<0,
\end{array}\right.  
\eeq
and $V(G_j)$ is given by Eq. (\ref{6.4.15}) with $G=G_j$.
\label{Th12}
\end{theorem}

{\bf Proof:} \begin{enumerate}
\item  For $k=0$, it is obvious.
For $k<0$, we have
\beqq
V_{k,x}-[U,V_k]&=&-\sum_{j=k}^{-1}\Big(V_{x}(G_j)-[U,V(G_j)]\Big)\lam^{2(k-j-1)}\\
&=& -\sum_{j=k}^{-1}U_*\left(K G_j-\lam^2 K G_{j-1}\right)\lam^{2(k-j-1)}\\
&=& U_*\left(\sum_{j=k}^{-1} K G_{j-1}\lam^{2(k-j)}-K
G_{j}\lam^{2(k-j-1)}\right) \\
&=& U_*\left( K G_{k-1}-K
G_{-1}\lam^{2k}\right) \\
&=& U_*( K G_{k-1})\\
&=& U_*(X_k).
\eeqq 

For the case of $k>0$, it is similar to prove.

\item  Noticing $U_{t_k}=U_*(u_{t_k})$, we obtain
\beqq
U_{t_k}-V_{k,x}+[U,V_k]= U_*(u_{t_k}-X_k).
\eeqq
The injectiveness of $U_*$ implies item 2 holds.

\end{enumerate}

From {\bf Theorem \ref{Th12}}, we immediately obtain the following corollary.
\begin{corollary}
The new hierarchy (\ref{mtk}) has Lax pair:
\beq
\psi_{xxx}&=&-\lam u \psi,\\
\psi_{t_k}&=&\sum\lam^{2(k-j)-1}\Big[ -6G_j\psi_{xx}
+3(G'_j+\lam\pa^{-3}\Upsilon G_j)\psi_x-
 (G''_j+3\lam\pa^{-2}\Upsilon G_j)\psi\Big], \nonumber\\
\eeq
where the related symbols are the same as Thereom \ref{Th12} and
Thereom
\ref{Th1}. 
\end{corollary}

So, all equations in the hierarchy (\ref{mtk}) have
 the Lax pair and are  therefore integrable.
 In particular, we have the following special cases.
\begin{itemize}
\item When we choose $G_{-1}=\frac{1}{6}$, 
equation (\ref{m-1})
has the following Lax pair:
\beq
\Psi_x&=&U(u,\lam)\Psi, \label{Ux}\\
\Psi_t&=&V(u,\lam)\Psi, \label{Vt}
\eeq
where $u=v_{xx}, $ $U(u,\lam) $ is defined by Eq. (\ref{U1}),
and $V(u,\lam)$ is given by
\beq
V(u,\lam)=\zhen{v_x}{-v}{\lam^{-1}}{0}{0}
{-v}{\lam vu}{0}{-v_x}.
\eeq
Apparently, Lax pair (\ref{Ux}) and (\ref{Vt}) 
is equivalent to
\beq
\psi_{xxx}&=&-\lam u \psi,\\
\psi_t&=&\lam^{-1}\psi_{xx}-v\psi_x+v_x\psi.
\eeq
which is a limit form in Ref. \cite{DHH[2002]} when $\al $
goes to $\infty$. 
\item In a similar way, choosing $G_0=u^{-\frac{2}{3}}$ gives
the Lax pair of  equation (\ref{m+2}), i.e. 
$ u_t=(u^{-\frac{2}{3}})_{xxxxx}$ 
\beq
\psi_{xxx}&=&-\lam u \psi,\\
\psi_t&=&-6\lam u^{-\frac{2}{3}} \psi_{xx}+3\lam
(u^{-\frac{2}{3}})_x\psi_x-\lam(u^{-\frac{2}{3}})_{xx}\psi.
\eeq
This Lax pair is different from/inequivalent to  the result in Ref. \cite{KD1}.
\item Furthermore, through choosing $G_0=\frac{(u^{-\frac{1}{3}})_{xx}
    -2(u^{-\frac{1}{6}})_{x}^2}{u}$, we  find that 
 the new equation 
(\ref{newequ1})
 has the Lax pair:
\beq
\psi_{xxx}&=&-\lam u \psi,\\
\psi_t&=&-6\lam G_0\psi_{xx}+3\lam
(G'_0+3\lam u^{-\frac{1}{3}})\psi_x-\lam(G''_0+9
\lam(u^{-\frac{1}{3}})_{xx})\psi.\nonumber\\
\eeq
\end{itemize}

\section{Nonlinearized $6N$-dimensional integrable system
from spectral problems}
To discuss the solution of the hierarchy (\ref{mtk}),
we use  the  constrained method which leads finite dimensional 
integrable system to the PDEs (\ref{mtk}).
Becasue Eq. (\ref{sp1})/(\ref{SP1}) is a 3rd order eigenvalue problem,
we have to investigate itself together with its adjoint problem when we adopt
the nonlinearized procedure \cite{C1}. Ma and Strampp \cite{MS1}
ever studied the AKNS and its its adjoint problem, a $2\times 2$ case,
by using the so-called symmetry constraint method. Now, we are
discussing a $3\times 3 $ problem related to the hierarchy
 (\ref{mtk}).

Let us return to  the spectral problem (\ref{SP1})
and consider its adjoint problem (\ref{sp2}), and change it to
the following matrix form
\beq
\Psi^*_x=\zhen{0}{0}{u\lam}{-1}{0}{0}{0}{-1}{0}\Psi^*, \ 
\Psi^*=\phlem{1}{2}{3}, \label{SP2}
\eeq
where $\psi^*=\psi^*_3 $.

Let $\lam_j\ (j=1,...,N)$ be $N$ distinct spectral values of (\ref{SP1}) and (\ref{SP2}),
and $q_{1j}, q_{2j}, q_{3j}$ and $p_{1j}, p_{2j}, p_{3j}$ be the corresponding
spectral functions, respectively. Then we have
\beq \begin{array}{l}
q_{1x}=q_2,\\
q_{2x}=q_3,\\
q_{3x}=-u\La q_1;
\end{array} \label{q+1}
\eeq
and 
\beq \begin{array}{l}
p_{1x}=u\La p_3,\\
p_{2x}=-p_1,\\
p_{3x}=-p_2,
\end{array} \label{p+1}
\eeq
where $\La=diag(\lam_1,...,\lam_N)$, 
$q_k=(q_{k1}, q_{k2},..., q_{kN})^T, \ p_k=(p_{k1}, p_{k2},..., p_{kN})^T, \ k=1,2,3.$

Let us consider the above systems  in the whole symplectic space
$(\R^{6N}, \ dp\wedge dq)$.    
We directly impose the following constraint:
\beq
u^{-\frac{2}{3}}=\sum^N_{j=1}\nabla\lam_j,\ \label{G0} 
\eeq
where $\nabla\lam_j=\lam_jq_{1j}p_{3j}$ is the functional gradient
of $\lam_j$ for spectral problems (\ref{SP1}) and (\ref{SP2}). Then Eq. (\ref{G0}) is saying
\beq
u=\left<\La q_1, p_3\right>^{-\frac{3}{2}} \label{m+}
\eeq
which composes a  constraint in the whole space $\R^{6N}$.
Under this constraint, Eq. (\ref{q+1}) and its
adjoint (\ref{p+1})
are cast in a Hamiltonial canonical form in $\R^{6N}$:
\beq \begin{array}{l}
q_x=\{q, H^+\},\\
p_x=\{p, H^+\}, \end{array}\label{pqH+}
\eeq
with the Hamiltonian
\beq
H^+=\left<q_2,p_1\right>+\left<q_3,p_2\right>+\frac{2}{
\sqrt{\left<\La q_1,p_3\right>}}, \label{H+}
\eeq
where $p=(p_1,p_2,p_3)^T, \ q=(q_1,q_2,q_3)^T \in \R^{6N}$,
, $\left<\cdot, \cdot\right>$ stands for the standard inner product in
$\R^N$, and we modify the usual Poisson bracket of two functions $F_1,F_2$ as follows:
\beq
\{F_1, F_2\}=\sum_{i=1}^3\left(\left<\frac{\pa F_1}{\pa q_i}, \frac{\pa F_2}{\pa p_i}\right>-\left<\frac{\pa F_1}{\pa p_i}, \frac{\pa F_2}{\pa q_i}\right>\right) \label{Poiss1}
\eeq  
which is still antisymmetric, bilinear and satifies the Jacobi identity.

To see the integrability of the system (\ref{pqH+}), we
take into account of the time part $\Psi_t=V_k\Psi $
and its adjoint $\Psi^*_t=-V^T_k\Psi^* $, where 
$V_k $ is defined by
$V_k=\sum_{j=0}^{k-1}
 V(G_j)\lam^{2(k-j-1)},$ and $V(G_j)$ is given by Eq. (\ref{6.4.15})
 with $G=G_j$.

Let us first look at $V_1$ case. Then the corresponding time part is:
\beq
\Psi_t=\lam\zhen{-(u^{-\frac{2}{3}})_{xx}}{3(u^{-\frac{2}{3}})_{x}}{-6u^{-\frac{2}{3}}}
{-(u^{-\frac{2}{3}})_{xxx}+6\lam u^{\frac{1}{3}}}
{2(u^{-\frac{2}{3}})_{xx}}{-3(u^{-\frac{2}{3}})_x}
{-(u^{-\frac{2}{3}})_{xxxx}}{(u^{-\frac{2}{3}})_{xxx}+6\lam
u^{\frac{1}{3}}}{-(u^{-\frac{2}{3}})_{xx}}\Psi, \label{SPt11}
\eeq
and its adjoint part is:
\beq
\Psi^*_t=\lam\zhen{(u^{-\frac{2}{3}})_{xx}}
{(u^{-\frac{2}{3}})_{xxx}+6\lam u^{\frac{1}{3}}}
{-(u^{-\frac{2}{3}})_{xxxx}}
{-3(u^{-\frac{2}{3}})_{x}}
{-2(u^{-\frac{2}{3}})_{xx}}{-(u^{-\frac{2}{3}})_{xxx}-6\lam
u^{\frac{1}{3}}}
{6u^{-\frac{2}{3}}}
{3(u^{-\frac{2}{3}})_x}
{(u^{-\frac{2}{3}})_{xx}}\Psi^*. \label{SPt12}
\eeq

Noticing the following relations
\beqq
u^{\frac{1}{3}}&=&\left<\La q_1, p_3\right>^{-\frac{1}{2}},\\
(u^{-\frac{2}{3}})_{x}&=&\left<\La q_2, p_3\right>
  -\left<\La q_1, p_2\right>,\\
(u^{-\frac{2}{3}})_{xx}&=&\left<\La q_3, p_3\right>
  +\left<\La q_1, p_1\right>-2\left<\La q_2, p_2\right>,\\
(u^{-\frac{2}{3}})_{xxx}&=&3\Big(\left<\La q_2, p_1\right>
  -\left<\La q_3, p_2\right>\Big),\\
(u^{-\frac{2}{3}})_{xxxx}&=&6\left<\La q_3, p_1\right>
+
3\left<\La q_1, p_3\right>^{-\frac{3}{2}}\Big(\left<\La^2 q_1, p_2\right>
  +\left<\La^2 q_2, p_3\right>\Big),
\eeqq
we obtain the nonlinearizations of the time parts (\ref{SPt11}) and
(\ref{SPt12}), and cast the nonlearized systems into
 canonical Hamiltonian system in $\R^{6N}$:
\beq \begin{array}{l}
q_{t_1}=\{q, F_1^+\},\\
p_{t_1}=\{p, F_1^+\}, \end{array}\label{pqF1+}
\eeq
with the Hamiltonian
\beq
F_1^+&=&-\frac{1}{2}\Big(\left<\La q_1,p_1\right>+\left<\La
  q_3,p_3\right>\Big)^2+2\left<\La q_2,p_2\right>\Big(\left<\La
  q_1,p_1\right>+\left<\La q_3,p_3\right>-\left<\La q_2,p_2\right>\Big)\nonumber\\
& & +3\Big(\left<\La q_2,p_3\right>-\left<\La q_1,p_2\right>\Big)
\Big(\left<\La q_2,p_1\right>-\left<\La q_3,p_2\right>\Big)-6
\left<\La q_1,p_3\right>\left<\La q_3,p_1\right>\nonumber\\
& & +\frac{6}{\sqrt{\left<\La q_1,p_3\right>}}
\Big(\left<\La^2 q_1,p_2\right>+\left<\La^2 q_2,p_3\right>\Big). \label{F1+}
\eeq
A direct computation leads to the following theorem.

\begin{theorem}

\beq 
\{H^+,F_1^+\}=0,
\eeq
that is, two Hamiltonian flows commute in $\R^{6N}$.
\end{theorem}

Furthermore, for general case $V_k, \ k>0, \ k\in \Z$, we consider the following
Hamiltonian functions
\beq
F_k^+&=&-\frac{1}{2}\sum_{j=0}^{k-1}
\Big(\left<\La^{2j+1} q_1,p_1\right>+\left<\La^{2j+1}
  q_3,p_3\right>\Big)\Big(\left<\La^{2(k-j)-1}
  q_1,p_1\right>+ \left<\La^{2(k-j)-1}
  q_3,p_3\right>\Big)\nonumber\\
& &+2\sum_{j=0}^{k-1}\left<\La^{2j+1} q_2,p_2\right>
\Big(\left<\La^{2(k-j)-1}
  q_1,p_1\right>+\left<\La^{2(k-j)-1}
  q_3,p_3\right>-\left<\La^{2(k-j)-1} q_2,p_2\right>\Big)\nonumber\\
& & +3\sum_{j=0}^{k-1}\Big(\left<\La^{2j+1} q_2,p_3\right>-\left<\La^{2j+1} q_1,p_2\right>\Big)
\Big(\left<\La^{2(k-j)-1} q_2,p_1\right>-\left<\La^{2(k-j)-1}
  q_3,p_2\right>\Big)\nonumber\\
& & -6\sum_{j=0}^{k-1}
\left<\La^{2j+1} q_1,p_3\right>\left<\La^{2(k-j)-1}
  q_3,p_1\right>\nonumber\\
& &-\frac{3}{2}\sum_{j=0}^{k}
\Big(\left<\La^{2j} q_1,p_1\right>-\left<\La^{2j}
  q_3,p_3\right>\Big)\Big(\left<\La^{2(k-j)}
  q_1,p_1\right>- \left<\La^{2(k-j)}
  q_3,p_3\right>\Big)\nonumber\\
& & -3\sum_{j=0}^{k}\Big(\left<\La^{2j} q_2,p_3\right>+\left<\La^{2j} q_1,p_2\right>\Big)
\Big(\left<\La^{2(k-j)} q_2,p_1\right>+\left<\La^{2(k-j)}
  q_3,p_2\right>\Big)\nonumber\\
& & +3H^+
\Big(\left<\La^{2k}
  q_1,p_2\right>+\left<\La^{2k}q_2,p_3\right>\Big). \label{Fk+}
\eeq

Then through a lengthy calculation, we find
\beq
\{H^+, F_k^+
\}=0, \{F^+_l, F_k^+
\}=0, \ k,l=1,2,...\ .
\eeq
That is,
\begin{theorem}
All canonical Hamiltonian flows 
$(F_k^+)$ commute with the Hamiltnonian system (\ref{pqH+}).
In particular, 
the Hamiltonian systems (\ref{pqH+}) 
and (\ref{pqF1+}) are compatible and therefore integrable in the Liouville sense.
\end{theorem}

\begin{remark}
 In the proof
of this Theorem, we use the following two facts:
$\left<q_1,p_2\right>+\left<q_2,p_3\right>=c_1,$
and $\left<q_1,p_1\right>-\left<q_3,p_3\right>=c_2$.
They  always hold along $x-$flow in the whole $\R^{6N}$.
Here $c_1,c_2 $ are two constants. 
\end{remark}

\begin{remark}
In fact, the involutive functions
$F_k^+ $ are generated from the nonlinearization
of the time part $\Psi_t=V_k\Psi $
and its adjoint part $\Psi^*_t=-V^T_k\Psi^* $ under 
the constraint (\ref{G0}), where 
$V_k $ is defined by
$V_k=\sum_{j=0}^{k-1}
 V(G_j)\lam^{2(k-j-1)},$ and $V(G_j)$ is given by Eq. (\ref{6.4.15})
 with $G=G_j$. In this calculation process, we use
the following equalities:
\beqq
G_j&=&-\left<\La^{2j+1}
      q_1,p_3\right>, \ j=0,1,2...\\
G_j'&=&\left<\La^{2j+1}
      q_2,p_3\right>-\left<\La^{2j+1}
      q_1,p_2\right>, \\
G_j''&=&\left<\La^{2j+1}
      q_3,p_3\right>+\left<\La^{2j+1}
      q_1,p_1\right>-2\left<\La^{2j+1}
      q_2,p_2\right>, \\
G_j'''&=&3\Big(\left<\La^{2j+1}
      q_2,p_1\right>-\left<\La^{2j+1}
      q_3,p_2\right>\Big),\\
G_j''''&=& 6\left<\La^{2j+1}
      q_3,p_1\right>+
3\left<\La q_1, p_3\right>^{-\frac{3}{2}}
\Big(\left<\La^{2j+2}
      q_1,p_2\right>+\left<\La^{2j+2}
      q_2,p_3\right>\Big),\\
\pa^{-1}mG_j'&=&\left<\La^{2j}
      q_3,p_2\right>+\left<\La^{2j}
      q_2,p_1\right>,\\
\pa^{-2}\Upsilon G_j&=& \left<\La^{2j}
      q_1,p_1\right>-\left<\La^{2j}
      q_3,p_3\right>, \\
\pa^{-3}\Upsilon G_j&=&-\Big( \left<\La^{2j}
      q_1,p_2\right>+\left<\La^{2j}
      q_2,p_3\right>\Big).
\eeqq
\end{remark}
\section{Parametric solution}

Since the Hamiltonian flows $(H^+)$ and $(F_k^+)$  are completely
integrable in $\R^{6N} $ and their Poisson brackets $\{H^+,F_k^+\}=0$ 
($k=1,2,...$),
their phase flows $g^x_{H^+},\ g^{t_k}_{F_k^+}$ commute \cite{AV}. Thus,
we can define their compatible solution as follows:
\beq
\left(\begin{array}{l}
q(x,t_k)\\
p(x,t_k)
\end{array}
\right)=g^x_{H^+} g^{t_k}_{F_k^+}\left(\begin{array}{l}
q(x^0,t_k^0)\\
p(x^0,t_k^0)
\end{array}
\right), \ \ k=1,2,...,
\eeq
where $x^0, \ t_k^0$ are the initial values of phase flows
$g^x_{H^+},\ g^{t_k}_{F_k^+}$.

\begin{theorem}
Let $q(x,t_k)=(q_1,q_2,q_3)^T, \ p(x,t_k)=(p_1,p_2,p_3)^T$ 
be the common solution of the two commutable  Hamiltonian flows
$(H_{^+})$ and $(F_k^+)$ in $\R^{6N}$. Then
\beq
u&=&\frac{1}
{\sqrt{\left<\La q_1(x,t_k), p_3(x,t_k)\right>^3}}, 
\label{m001}
\eeq
satisfies the positive order hierarchy
   \beq
u_{t_k} & = & J{\cal L}^{k}\cdot u^{-\frac{2}{3}},
   \ \  k=1,2,...,
   \label{unew+}
   \eeq
where the operators ${\cal L}=J^{-1}K$, $J,\ K$
 are given by Eqs. (\ref{J2})
and (\ref{K2}), respectively.
   \label{th4+}
\end{theorem}

{\bf Proof}: Direct computation completes this proof.

\begin{theorem}
Let $p(x,t),q(x,t)$ ($p(x,t)=(p_1, p_2,p_3)^T,  q(x,t)=(q_1,
q_2,q_3)^T$) be the common solution of the two integrable 
commutable  flows
(\ref{pqH+}) and (\ref{pqF1+}), then 
\beq
u&=&\frac{1}
{\sqrt{\left<\La q_1(x,t), p_3(x,t)\right>^3}}, \label{m01}
\eeq
satisfies the equation:
\beq
u_t=\pa^5_x u^{-\frac{2}{3}}.
\eeq
\end{theorem}

{\bf Proof:} \ \ Doing five times derivatives in $x$ on
both sides of Eq. (\ref{m01}), we obtain
 \beq
\pa^5_x u^{-\frac{2}{3}}
=9u\Big(\left<\La^2 q_3, p_3\right>-\left<\La^2 q_1,
  p_1\right>\Big)+3u_x\Big(\left<\La^2 q_1, p_2\right>+\left<\La^2 q_2,
  p_3\right>\Big), \label{m5}
\eeq
where
\beqq
u_x=-\frac{3}{2}u\frac{\Big(\left<\La^2 q_1, p_2\right>+\left<\La^2 q_2,
  p_3\right>\Big)\Big(\left<\La q_2, p_3\right>-\left<\La q_1,
  p_2\right>\Big)
}{\left<\La q_1, p_3\right>}.
\eeqq

On the other hand, doing the derivative in $t$ on
the both sides of Eq. (\ref{m01}) yields
\beqq
u_t&=&-\frac{3}{2}u\frac{\left<\La p_3, \dot{q}_1\right>+\left<\La q_1,
  \dot{p}_3\right>
}{\left<\La q_1, p_3\right>}\\
&=&-\frac{3}{2}u\frac{\left<\La p_3, \frac{\pa F_1^+}{\pa p_1}\right>-\left<\La q_1,
 \frac{\pa F_1^+}{\pa q_3}\right>
}{\left<\La q_1, p_3\right>}. 
\eeqq
Substituting the expression of $F_1^+$ into the above
equality and calculating, we find that this final result
is the same as  the right hand side of Eq. (\ref{m5}),
which completes the proof.

\section{Traveling wave solutions}

{\bf First},  Let us  compute traveling wave solution for equation 
(\ref{newequ2}).  Set $u=f(\xi), \ \xi=x-ct$ ($c$ is some constant
speed), then after substituting this setting into
 equation (\ref{newequ2})
we obtain
\beqq
-cf'''+3f''f'+f'''f=0,
\eeqq
i.e.
\beqq
(f^2-2cf)'''=0.
\eeqq
Therefore, 
\beq
(f-c)^2=A\xi^2+B\xi+C, \ \ \forall \  A,\ B, \ C\in \R.
\eeq
So, the equation $u_{xxt}+3u_{xx}u_x+u_{xxx}u=0$ has the following
traveling wave solution
\beq
u(x,t)=c\pm\sqrt{A(x-ct)^2+B(x-ct)+C}. \label{u-newequ2}
\eeq

Let us discuss several special cases:
\begin{itemize}

\item When $c=0$, we get stationary solution
\beq
u(x)=\pm\sqrt{Ax^2+Bx+C}, \ \forall A,\ B, \ C\in \R,
\eeq 
which can be a straight line, circle, ellipse, parabola, and hyperbola
according to different choices of constants $A, \ B, C.$

\item When $c\not=0$ and $A\not=0$, then we have
\beq
u(x,t)=c\pm\sqrt{A\Big(x-ct+\frac{B}{2A}\Big)^2+\frac{4AC-B^2}{4A}}, \ \forall A,\ B, \ C\in \R,
\eeq 
therefore if $4AC-B^2=0$ this solution becomes
\beq
u(x,t)=c\pm\sqrt{A}\Big|\Big(x-ct+\frac{B}{2A}\Big)\Big|, \ \forall A>0, \ B \in \R
\eeq
 For example, setting $c=1,\ A=1, \ B=0 $
yields
\beq
u(x,t)=
1-|x-t|,
\eeq
and
\beq
u(x,t)=1+|x-t|.
\eeq
The former looks like a compacton solution  \cite{Rosenau,Fringer}.
The latter is a ``V"-type solution. 


\item When $c\not=0$ and $A=0$, then we have
\beq
u(x,t)=c\pm\sqrt{B(x-ct)+C}, \
\forall \ B, \ C\in \R,
\eeq
which is a parabolic traveling wave solution  if $B\not=0$
and becomes a constant solution if $B=0$.
For example, the following  
\beq
u(x,t)=1+\sqrt{x-t}, x-t\geq0,
\eeq
and
\beq
u(x,t)=1-\sqrt{x-t}, x-t\geq0.
\eeq 
are two special solutions.
\end{itemize}

So, the 3rd-order equation $u_{xxt}+3u_{xx}u_x+u_{xxx}u=0$  has the continuous traveling
wave solution
(\ref{u-newequ2}).  

\begin{figure}[ht!]
\centerline{
\scalebox{0.60}{
\includegraphics{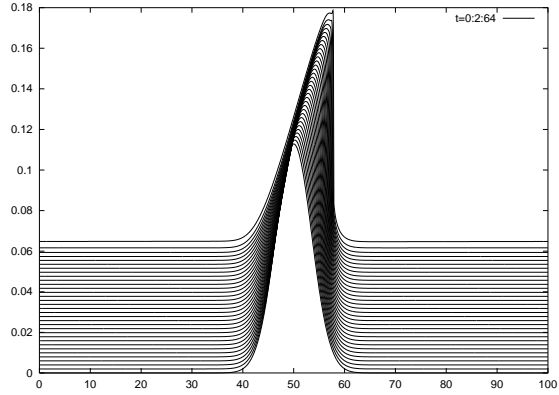}
}
}
\caption{Stable  period from $t=0 $ to $t=64$  for the equation
  $u_{xxt}+3u_{xx}u_x+u_{xxx}u=0$ 
under the Gaussian initial condition. This figure is very like the Burgers
case 
$u_{xxt}+3u_{xx}u_x+u_{xxx}u-\epsilon u_{xxxx}=0$
which is formed through adding small viscosity term $\epsilon u_{xxxx}$ to the
equation. For instance, when $\epsilon=0.01 $, the equation
 has Figure
\ref{3rd-order-with-Burgers}.}
\label{qiao5-3rd-sln-fig}
\end{figure}

The  Gaussian initial solution of this 3rd-order PDE
is stable from $t=0$ to $t=64$
(see Figure \ref{qiao5-3rd-sln-fig}).But after $t=64$
this solution is not satble (see Figures 2 - 5).


\begin{figure}[ht!]
\centerline{
\scalebox{0.60}{
\includegraphics{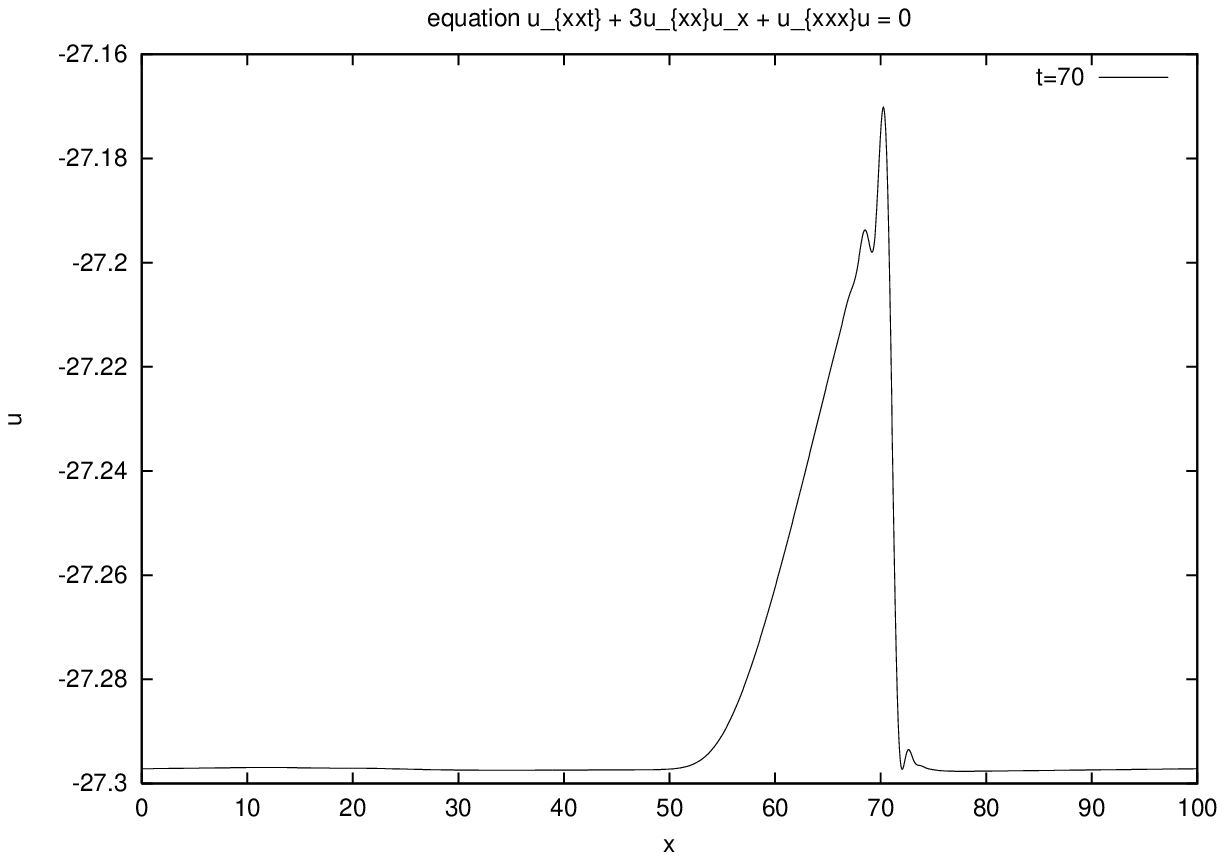}
}
}
\caption{Solution shape at $t=70$  for the equation
  $u_{xxt}+3u_{xx}u_x+u_{xxx}u=0$.}
\label{qiao1-fig}
\end{figure}

\begin{figure}[ht!]
\centerline{
\scalebox{0.60}{
\includegraphics{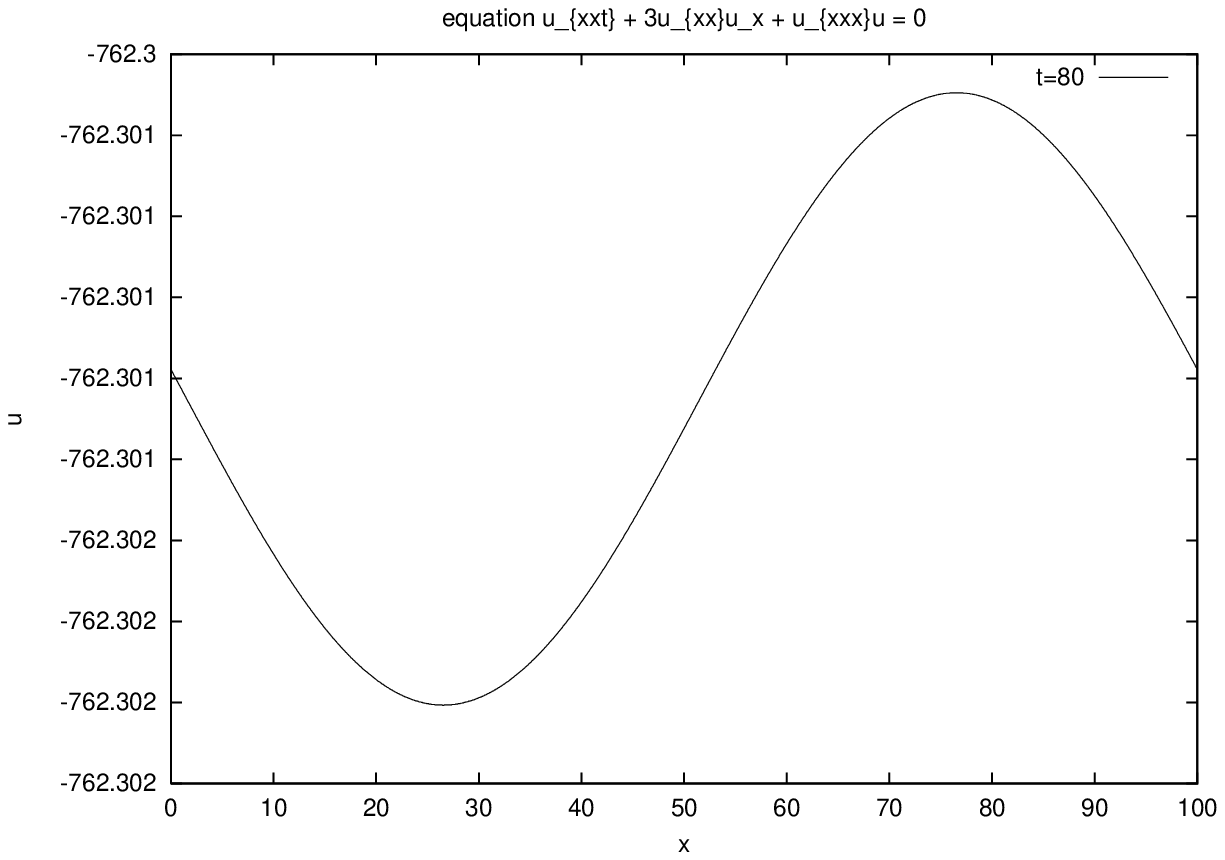}
}
}
\caption{Solution shape at $t=80$  for the equation
  $u_{xxt}+3u_{xx}u_x+u_{xxx}u=0$.}
\label{qiao2-fig}
\end{figure}

\begin{figure}[ht!]
\centerline{
\scalebox{0.60}{
\includegraphics{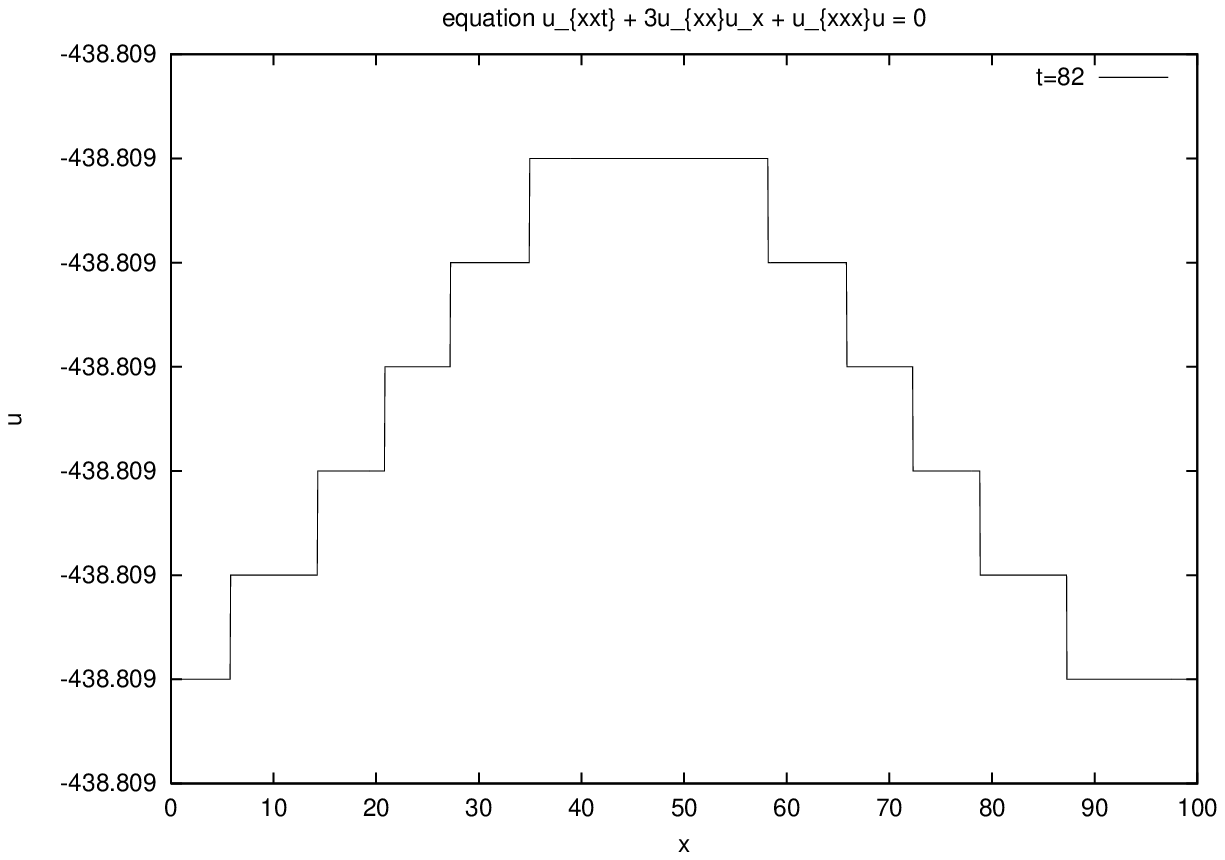}
}
}
\caption{Solution shape at $t=82$  for the equation
  $u_{xxt}+3u_{xx}u_x+u_{xxx}u=0$.}
\label{qiao3-fig}
\end{figure}

\begin{figure}[ht!]
\centerline{
\scalebox{0.60}{
\includegraphics{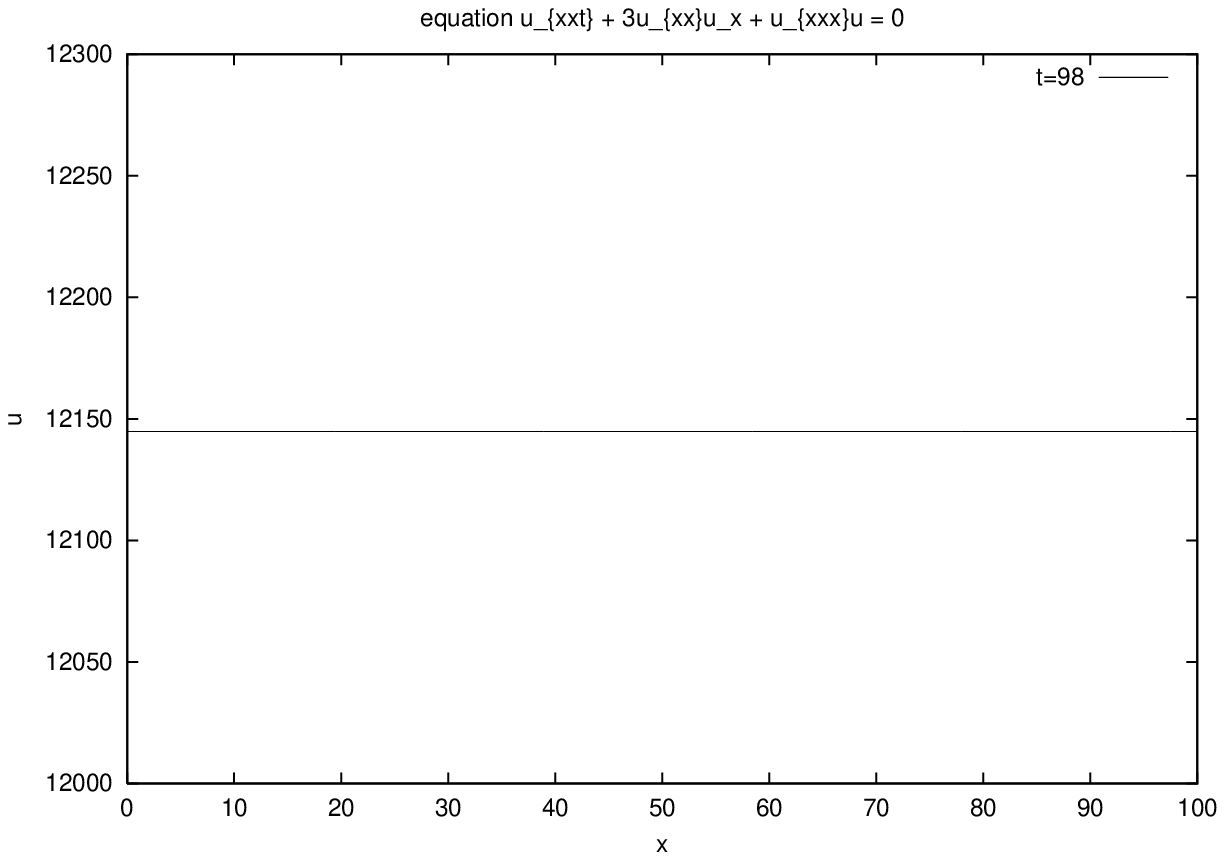}
}
}
\caption{Solution shape at $t=98$  for the equation
  $u_{xxt}+3u_{xx}u_x+u_{xxx}u=0$.}
\label{qiao4-fig}
\end{figure}

\begin{figure}[ht!]
\centerline{
\scalebox{0.60}{
\includegraphics{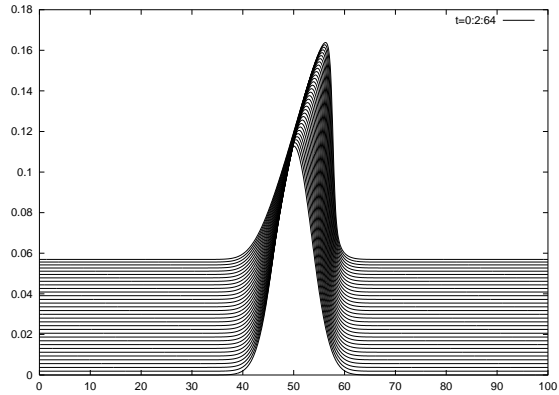}
}
}
\caption{Stable  solution  for the equation
$u_{xxt}+3u_{xx}u_x+u_{xxx}u-\epsilon u_{xxxx}=0, \ \epsilon=0.01$  
under the Gaussian initial condition. This figure is almost same as
Figure \ref{qiao5-3rd-sln-fig}. But, for large $\epsilon$ they are
quite different, and for negative  $\epsilon$, the corresponding solution  blows up.
}
\label{3rd-order-with-Burgers}
\end{figure}

$ $


\newpage

{\bf Second}, we give  traveling wave solution for the 5th-order equation 
(\ref{newequ}).  Set $u=\xi^{-\gamma}, \ \xi=x-ct$ ($c$ is some constant
speed to be determined), then after substituting this setting into
 equation (\ref{newequ})
we obtain
\beq
\gamma=\frac{12}{5}, \ \ c=-\frac{336}{625}.
\eeq
So, the 5th-order equation 
(\ref{newequ}) has the following traveling wave solution 
\beq
u=(x+\frac{336}{625}t)^{-\frac{12}{5}}. \label{u-5th}
\eeq

\begin{figure}[ht!]
\centerline{
\scalebox{0.80}{
\includegraphics{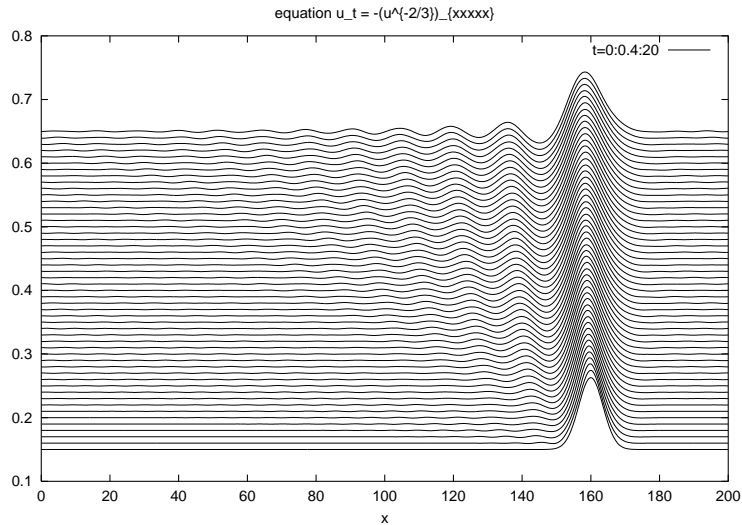}
}
}
\caption{This is the stable solution for the 5th-order equation
  $u_t=\pa^5_x u^{-2/3} $ under the Gaussian initial
  condition.}
\label{5th-Gaussian-sln-fig}
\end{figure}

Although at each time the solution (\ref{u-5th}) has
singular point at $x=-\frac{336}{625}t$, this 5th-order PDE
has the smooth and stable traveling wave solution under the Gaussian initial condition
(see Figure \ref{5th-Gaussian-sln-fig}). 

\begin{figure}[ht!]
\centerline{
\scalebox{0.80}{
\includegraphics{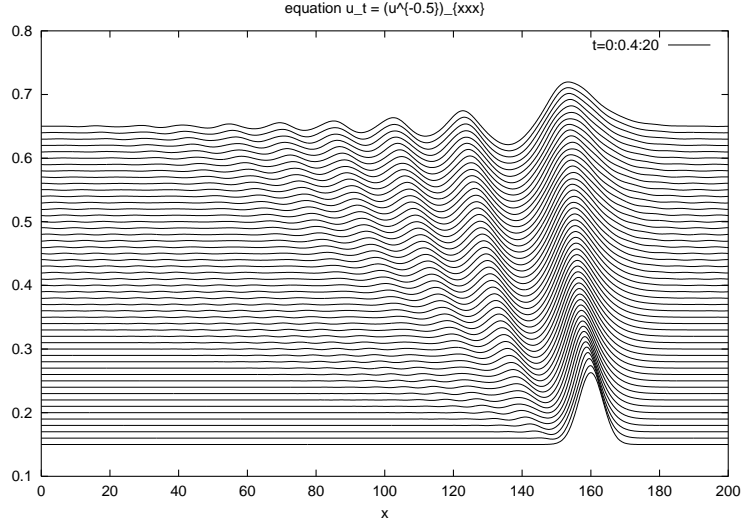}
}
}
\caption{This is the stable solution for the HD equation
  $u_t=\pa^3_x u^{-1/2} $ under the Gaussian initial
  condition.}
\label{3rd-Gaussian-sln-fig}
\end{figure}

\newpage

So, the figure  \ref{5th-Gaussian-sln-fig}
of the equation $u_t=\pa^5_x u^{-2/3} $
has a slight difference from the figure \ref{3rd-Gaussian-sln-fig}
of the HD equation 
$u_t=\pa^3_x u^{-1/2} $.


$ $

{\bf Third}, we give  traveling wave solution for the new integrable
7th-order equation 
(\ref{newequ1}).  Set $u=\xi^{-\gamma}, \ \xi=x-ct$ ($c$ is some constant
speed to be determined), then 
we have
\beq
\gamma=\frac{18}{7}, \ \ c=\frac{31680}{117649}.
\eeq
So, the 7th-order equation 
(\ref{newequ1}) has the following traveling wave solution
\beq
u=(x-\frac{31680}{117649}t)^{-\frac{18}{7}}.
\eeq



$ $

Furthermore, we propose the following new equation:
\beq
u_t=\partial_x^l u^{-m/n}, \ l\ge1,\ n\not=0,\ \ m,n \in \Z. \label{utmn}
\eeq
This equation has the following  traveling wave solution
\beq
u(x,t)&=&(x-ct)^{-n(l-1)/(m+n)},\label{uxct}\\
& & c=\frac{m}{n}\prod_{k=1}^{l-1}\Big(\frac{m(l-1)}{m+n}-k\Big). \nonumber
\eeq

Apparently, if $mn+n^2>0$ this solution has singularity at $x_0=ct_0 $
 ($t_0$ is some time), and 
if $mn+n^2<0$ this solution is a smooth  traveling wave  solution.


\begin{remark}
Here are the special cusp-like traveling wave solutions
\beq
u(x,t)=(x-\frac{2}{9}t)^{-4/3}
\eeq
and
\beq
u(x,t)=(x+\frac{336}{625}t)^{-12/5}
\eeq
for the Harry-Dym equation
$
u_t=\pa^3(u^{-1/2})
$
and
the 5th order equation $
u_t=\pa^5(u^{-2/3}).
$

\end{remark}

\section{Conclusions}

In section 5, we obtain a parametric solution
(\ref{m01}) of the 5th-order equation (\ref{newequ}).
This parametric solution can not include its traveling
wave solution $u=(x+\frac{336}{625}t)^{-\frac{12}{5}}$
because the constrained relation
\beqq
\left<\La q_1(x,t), p_3(x,t)\right>=\Big(x+\frac{336}{625}t\Big)^{\frac{8}{5}}
\eeqq  
does not hold, where 
\beqq
\pa^3_x q_1&=&-\Big(x+\frac{336}{625}t\Big)^{-\frac{12}{5}}\La q_1,\\
\pa^3_x p_3&=&\Big(x+\frac{336}{625}t\Big)^{-\frac{12}{5}}\La p_3.
\eeqq

Traveling wave solutions $u=(x+\frac{336}{625}t)^{-\frac{12}{5}}$
for equation $u_t=\pa^5u^{-2/3}$ and $u=(x-\frac{31680}{117649}t)^{-\frac{18}{7}}$
for equation $u_t=\pa^5_x\frac{(u^{-\frac{1}{3}})_{xx}
    -2(u^{-\frac{1}{6}})_{x}^2}{u}$
are singular at some certain ponits $x$ with the different  time $t$.
That is, this singularity travels with the time $t$.  In the case of  $n(m+n)>0$, the traveling
wave solution (\ref{uxct}) for general equation (\ref{utmn})
is also matching this case. 

A natural question arises here:  is the equation $u_t=\partial_x^l u^{-m/n}$
 integrable for all $l\ge1, m,n,\in \Z $ or for what kind of
$l\ge1, m,n,\in \Z $ it is integrable? So far, for the cases: $l=3, m=1, n=2 $ and 
$l=5, m=2, n=3 $, we know that it is integrable.

 The Harry-Dym equation has the cusp-like traveling wave solution
$u(x,t)=(x-\frac{2}{9}t)^{-4/3}$, but this is not cusp soliton
which Wadati described this in Ref. \cite{Wadati}, because the traveling wave solution
is singular, but the cusp is continuous. 

If we consider other constraints between the potential
and the eigenfunctions, then we can still get the parametric solutions
for other two equations (\ref{newequ1}) and (\ref{newequ2}).

\subsection*{Acknowledgments}
We thank
Prof. Konopelchenko for reminding his paper \cite{KD1} and Prof. Magri
for his fruitful discussion during their visit at Los Alamos National
Laboratory. We  also thank Dr. Shengtai Li for his  help in drawing
  Figures.  

This work was supported by the U.S. Department of Energy under
contracts W-7405-ENG-36 and the Applied Mathematical Sciences Program
KC-07-01-01; and also the Special Grant of National
Excellent Doctorial Dissertation of PR China.


\begin{thebibliography}{99}
\parskip=0.01cm

\bibitem{AKNS}Ablowitz M J,  Kaup D J,  Newell A C,  Segur H,
             Nolinear evolution equations of physical significance,
             {\it Phys. Rev. Lett.}  31(1973), 125-127.

\bibitem{AKNS1}Ablowitz M J,  Kaup D J,  Newell A C,  Segur H,
  {\it Studies in Appl. Math. } 53(1974), 249-315.
\bibitem{AV}V. I. Arnol'd, {\it Mathematical Methods of Classical Mechanics}
            (Springer-Verlag, Berlin, 1978).
\bibitem{CH1}Camassa R, Holm D D,
          An integrable shallow water
          equation with peaked solitons,
                     {\it Phys. Rev. Lett.}  71 (1993), 1661-1664.
\bibitem{C1}Cao C W,
Nonlinearization of Lax system
             for the AKNS hierarchy,  {\it Sci. China A} (in Chinese) 32(1989), 701-707;
             also see English Edition:
              Nonlinearization of Lax system for the AKNS hierarchy,
              {\it Sci. Sin. A} 33(1990), 528-536.
\bibitem{DHH[2002]}
Degasperis A, Holm D D,  Hone A N W,
A new integrable equation with peakon
solutions, {\it NEEDS (2002) Proceedings}, to appear.
\bibitem{DP[1999]}
Degasperis A and Procesi M, Asymptotic integrability, in Symmetry and Perturbation Theory, edited by A. Degasperis and G. Gaeta, World Scientific (1999) pp.23-37.

\bibitem{Fringer}Fringer D,  Holm D D,
Integrable vs. nonintegrable geodesic soliton behavior,
{\it Physica D} 150(2001), 237-263.

\bibitem{GGKM}Gardner C S, Greene J M, Kruskal M D, Miura R M,
         Method for Solving the Korteweg-de Vries Equation, {\it Phys. Rev. Lett.}
         19(1967), 1095-1097.

\bibitem{HH[2002]}Holm D D,  Hone A N W, Note on Peakon Bracket,
Private communication, 2002.
\bibitem{Holm-Qiao-2002}Holm D D,  Qiao Z, Integrable hierarchy,
  $3\times3 $ constrained systems, and parametric and peaked
  stationary
solutions, preprint, 2002.


\bibitem{Kaup1}Kaup D J, On the inverse scattering problem for
         cubis eigenvalue problems of the class
         $\psi_{xxx}+6Q\psi_x+6R\psi=\lam\psi$, {\it
         Stud. Appl. Math.} 62(1980), 189-216.
\bibitem{KD1}Konopelchenko B G,  Dubrovsky V G, Some new
integrable nonlinear evolution equations in $2+1$ dimensions,
 {\it Phys. Lett. A} 102(1984), 15-17.
\bibitem{Kup1}Kuperschmidt B A, A super Korteweg-De Vries
         equation: an integrable system,
 {\it Phys. Lett. A} 102(1984), 213-215.
 
\bibitem{KDV}Korteweg D J,  Vries De G, On the change of form
  long waves advancing in a rectangular canal, and on a new type of
  long stationary waves, {\it Phil. Mag.} 39(1895), 422-443.
\bibitem{LG}Levitan B M, Gasymov M G, Determination of a differential
  equation by two of its spectra, {\it Russ. Math. Surveys} 19:2(1964), 1-63. 
\bibitem{MS1}Ma W X, Strampp W, An explicit symmetry constraint for the Lax pairs and the adjoint Lax pairs of AKNS systems, 
{\it Phys. Lett. A} 185(1994), 277-286. 

\bibitem{Mar}Marchenko V A,  Certain problems in the theory of
  second-order differential operators, {\it Doklady Akad. Nauk SSSR} 72(1950), 457--460 (Russian).

\bibitem{Qiao} Qiao Z J,  {\it Finite-dimensional Integrable System and Nonlinear Evolution Equations}, Higher Education
  Press, PR China, 2002.

\bibitem{Rosenau}Rosenau P, Hyman J M, Compactons: Solitons with
  finite wavelength, {\it Phys. Rev. Lett} 70(1993), 564 -- 567.

\bibitem{TuGZ}Tu G Z,  An extension of a theorem on gradients of conserved
            densities of integrable systems, {\it Northeast. Math. J.} 6(1990),
            26-32.
\bibitem{Wadati}Wadati M,  Ichikawa Y H, Shimizu T, Cusp soliton of a new integrable nonlinear
evolution equation, 
{\it Prog. Theor. Phys.} 64(1980), 1959-1967.
\bibitem{ZS}Zakharov V E,  Shabat A B, Exact theory of two dimensional
         self focusing and one dimensional self modulation of waves in nonlinear media,
          {\it Sov. Phys. JETP} 34(1972), 62-69.
\end{thebibliography}
\end{document}